\documentclass[prb,prl,10pt,oneside,twocolumn]{revtex4}

\usepackage{graphicx}
\usepackage{dcolumn}
\usepackage{epstopdf}
\usepackage{bm}
\usepackage{amssymb}
\usepackage{amsmath}
\usepackage{parskip}
\usepackage{multirow}
\usepackage{dcolumn}
\usepackage{hyperref} 
\usepackage[sort&compress]{natbib}

\setcitestyle{square}

\begin{document}

\title{Ti$_{1-x}$Au$_x$ Alloys: Hard Biocompatible Metals and Their Possible Applications}
\author{E. Svanidze,$^1$ T. Besara,$^2$ M. F. Ozaydin,$^3$ Yan Xin,$^2$ K. Han,$^2$ H. Liang,$^3$ T. Siegrist,$^2$ and E. Morosan$^1$}

\address{$^1$Department of Physics and Astronomy, Rice University, Houston, TX 77005, USA}
\address{$^2$National High Magnetic Field Laboratory, Florida State University, Tallahassee, FL 32306, USA}
\address{$^3$Department of Mechanical Engineering, Texas A\&M University, College Station, TX 77843, USA}

\date{\today}

\begin{abstract}

The search for new hard materials is often challenging from both theoretical and experimental points of view. Furthermore, using materials for biomedical applications calls for alloys with high biocompatibility which are even more sparse. The Ti$_{1-x}$Au$_x$ ($0.22 \leq x \leq 0.8$) exhibit extreme hardness and strength values, elevated melting temperatures (compared to those of constituent elements), reduced density compared to Au, high malleability, bulk metallicity, high biocompatibility, low wear, reduced friction, potentially high radio opacity, as well as osseointegration. All these properties render the Ti$_{1-x}$Au$_x$ alloys particularly useful for orthopedic, dental, and prosthetic applications, where they could be used as both permanent and temporary components. Additionally, the ability of Ti$_{1-x}$Au$_x$ alloys to adhere to ceramic parts could reduce the weight and cost of these components. 

\end{abstract}

\maketitle

\section{I. Introduction}

In addition to numerous applications in the industrial, automotive and aerospace fields, Ti has been widely used for implant devices that replace patients' hard tissues  \cite{Elias_2008, Semlitsch_1987}. A number of \textit{in vivo} and \textit{in vitro} experiments with various grades of Ti concluded that commercially pure Ti is a highly biocompatible material due to the spontaneous build-up of an inert and stable oxide layer \cite{Elias_2008, Johansson_1991}. Additional properties that make Ti suitable for biomedical applications include high strength-to-weight ratio \cite{ Hautaniemi_1992, Wang_1996}, low electrical conductivity, low ion-formation levels in aqueous environments, low pH value, and a dielectric constant comparable to that of water \cite{Elias_2008}. Moreover, Ti is one of a few materials capable of osseointegration -- mechanical retention of the implant by the host bone tissue -- which stabilizes the implant without any soft tissue layers between the two \cite{Branemark_1983}. These properties enable a wide use of Ti for devices such as artificial knee and hip joints, screws and shunts for fracture fixation, bone plates, pacemakers and cardiac valve prosthesis \cite{Samsonov_1968,Marya_2010}. Not surprisingly, the dental applications of Ti are just as common, including implants and their components such as inlays, crowns, overdentures, and bridges \cite{Elias_2008, Kempf_1996, Kempf_1996_2,Cascone_2011, Vuilleme_2003, Takahashi_2004}. 

However, the pure Ti is not strong enough for a number of medical purposes \cite{Ho_2008, Faria_2011}, calling for the development of more superior Ti-based alloys \cite{Okuno_1989, Lautenschlager_1993, Hattori_2001, Takada_2001, Takahashi_2002}. Additionally, Ti exhibits poor machineability, which reduces tool life, increases the processing time and is problematic when the elimination of a dental Ti prosthesis is necessary \cite{Nakajima_1996, Okuno_1989}. Both the machineability and hardness can be improved by alloying Ti with another element. After a number of toxic effects were reported in permanent implants \cite{Elias_2008}, the use of V- and Al-containing Ti alloys was discontinued. Among biocompatible elements, the addition of Ag and Cu nearly doubles the hardness, compared to pure Ti \cite{Kikuchi_2003, Takahashi_2002, Kikuchi_2003b, Kikuchi_2003c}. Biocompatible Au, which is located in the same group as Ag and Cu, is a $\beta$-stabilizer -- Ti occurs in two different crystallographic structures (hexagonal \textit{P6$_3$/mmc} $\alpha$-Ti and higher hardness cubic \textit{Im$\overline{3}$m} $\beta$-Ti). Since both Ag and Cu are $\alpha$-stabilizers, $\alpha$-Ti was produced by alloying Ti with these metals. It can be expected that, if it were possible to form the equivalent $\beta$-Ti-Au alloys, their hardness may increase. Moreover, the high biocompatibility and corrosion resistance of Au may yield an alloy suitable for biomedical purposes \cite{Kempf_1996_2}, given the wide use of Au and Au-doped implant devices \cite{Fischer_2000, Fischer_2000, Gafner_1989, Humpston_1992, Takahashi_1998}. In case the machineability decreases with increased hardness, the relatively low melting temperatures of Ti-Au alloys will allow for the majority of parts to be used in an as-cast form. Additionally, the Ti-Au alloys can adhere to a ceramic surface, making it convenient for a number of biomedical applications, reducing the overall weight and cost of the corresponding parts \cite{Fischer_2000, Strietzel_2009}.

In this manuscript we present the hardness, wear rate as well as the microstructural analysis of Ti$_{1-x}$Au$_x$ ($0.22 \leq x \leq 0.8$) alloys indicating that the $x = 0.25$ alloy is suitable for a number of biomedical applications, particularly where Ti is already employed. Despite the fact that Au is a $\beta$-Ti stabilizer, it has been observed that the $x = 0.25$ alloy forms two microstructures: $\alpha$Ti and Ti$_3$Au. A remarkable nearly four-fold increase in hardness, as compared with pure Ti, is registered for the Ti$_{0.75}$Au$_{0.25}$ alloy. The hardness values exceed those of most biocompatible alloys and engineering metals, as well as most engineering ceramics.

\section{II. Experimental Methods}

Alloys of Ti$_{1-x}$Au$_x$ were prepared by arcmelting Ti (Cerac, 99.99\%) and Au (Cerac, 99.99\%) in stoichiometric ratios, with mass losses no more than 0.3 \%. To ensure homogeneity, the samples were re-melted several times.

The Vickers hardness $H_V$ was estimated using a Tukon 2100 microhardness tester, equipped with a Vickers diamond pyramid indenter. The microhardness tests were performed on a polished sample surface of about 3 mm in diameter. Multiple tests were conducted for all samples, using a 300 g load, with a duration of 10 s. Another important quantity is the yield strength $\sigma_f$ \cite{Ashby_2011}, which, for metals, is defined as the stress at which the stress-strain curve for axial load deviates by a strain of 0.2 $\%$ from the linear elastic line \cite{Ashby_2011}. At times when stress-strain data are not available, as is the case in Ti$_{1-x}$Au$_x$ alloys, the value of $\sigma_f$ can be extracted from the Vickers hardness $H_V$ using $\sigma_f = H_V/3$ \cite{Ashby_2011}. The $H_V$ is the Vickers hardness in units of MPa, resulting in the same units for $\sigma_f$. This equation has been empirically shown to be approximately true for metals \cite{Tabor_1951}. 

\section{III. Data Analysis and Discussion}

Although both the Ti-rich and Au-rich sides of the Ti-Au phase diagram were previously explored in detail, hardly anything has been reported on the hardness of intermediate compositions. In order to further investigate this system, a series of Ti$_{1-x}$Au$_x$ alloys was prepared with $0.22 \leq x \leq 0.8$.

\begin{figure}
\centering
\includegraphics [width=\columnwidth] {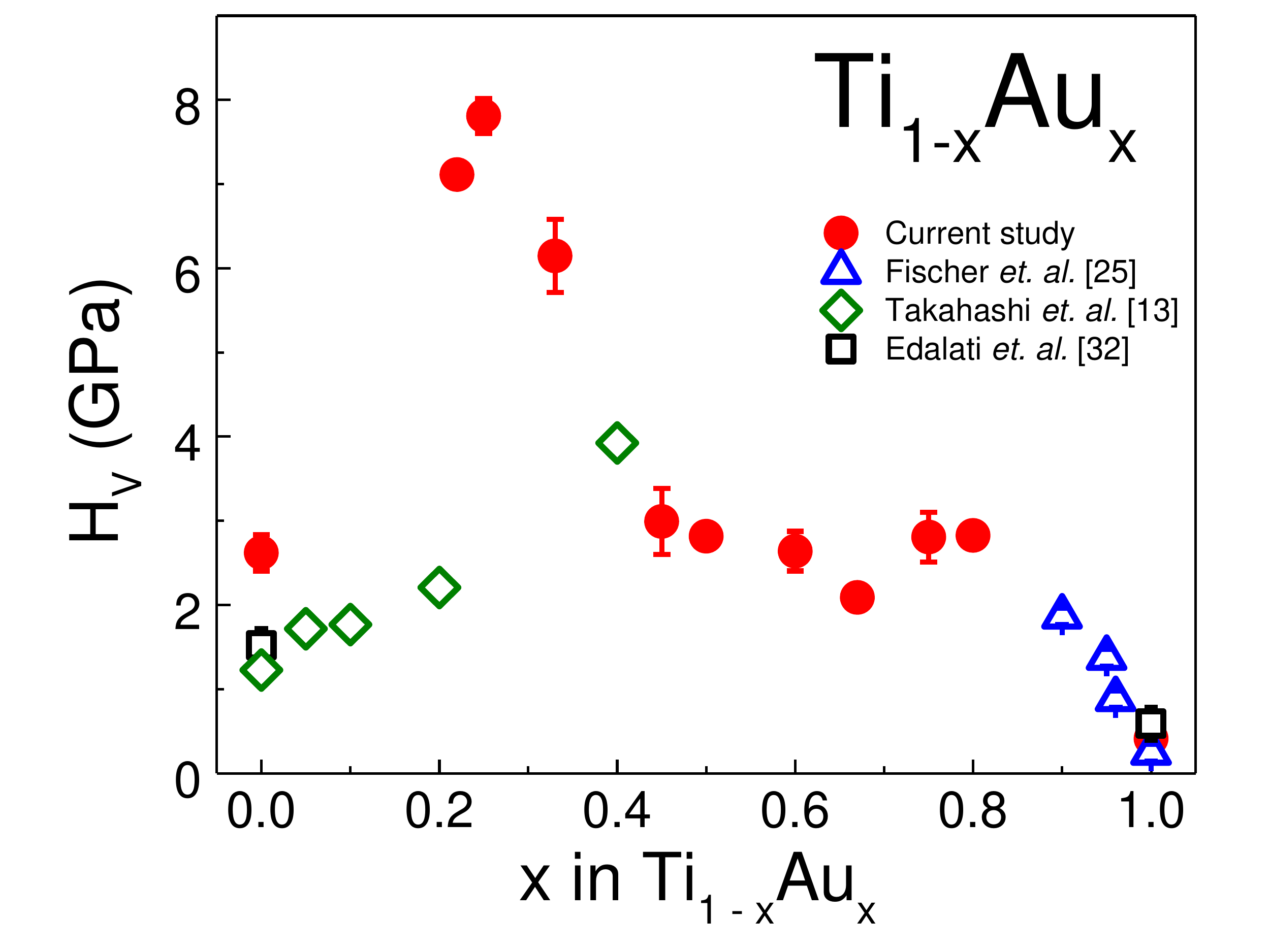}
\caption{Hardness $H_V$ as a function of composition for Ti$_{1-x}$Au$_x$ alloys: current study (circles), Au-rich (triangles) \cite{Fischer_2000} and Ti-rich (diamonds) \cite{Takahashi_2004} regions, along with elemental data (squares) \cite{Edalati_2010}.}
\label{Hardness}
\end{figure}

\begin{figure}
\centering
\includegraphics [width=\columnwidth] {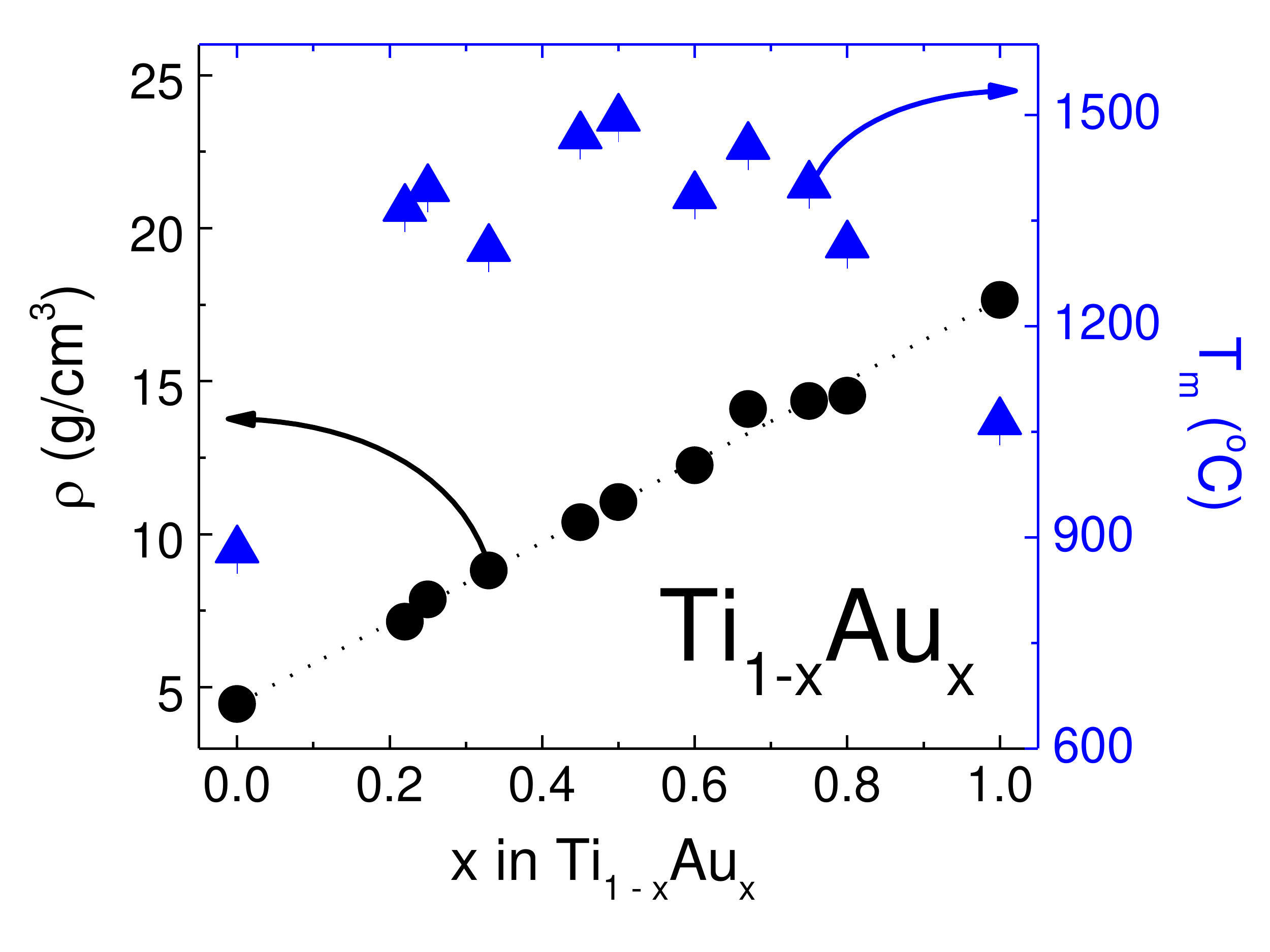}
\caption{Mass density $\rho$ (left axis, circles) and melting temperature $T_m$ \cite{ASM_2007} (right axis, triangles) as a function of composition for Ti$_{1-x}$Au$_x$ alloys ($0 \leq x \leq 1$).}
\label{Density}
\end{figure}

Surprisingly, the measurements reveal a non-monotonous change of hardness with $x$. Even more remarkable is the broad peak in $x$ between 0.22 and 0.35, where the hardness is up to three or four times higher than that of pure Ti. This indicates that the Ti$_{1-x}$Au$_x$ alloys with $0.22 \leq x \leq 0.35$ are likely better suited for applications where Ti is currently used \cite{Samsonov_1968, Marya_2010, Elias_2008, Kempf_1996, Kempf_1996_2,Cascone_2011, Vuilleme_2003}. Of the measured compositions, the $x = 0.25$  alloy exhibits the highest hardness value of 7.81 GPa, even higher than that of pearlitic steels \cite{Han_2001, Raabe_2010} and similar to that of high-carbon steels \cite{Cui_2007}. As mentioned above, both Ti and Au are biocompatible and have high resistance to \textit{in vivo} corrosion, suggesting that the resulting alloys are suitable for biomedical applications \cite{Elias_2008, Johansson_1991, Kempf_1996_2}.

\begin{figure}
\centering
\includegraphics [width=\columnwidth] {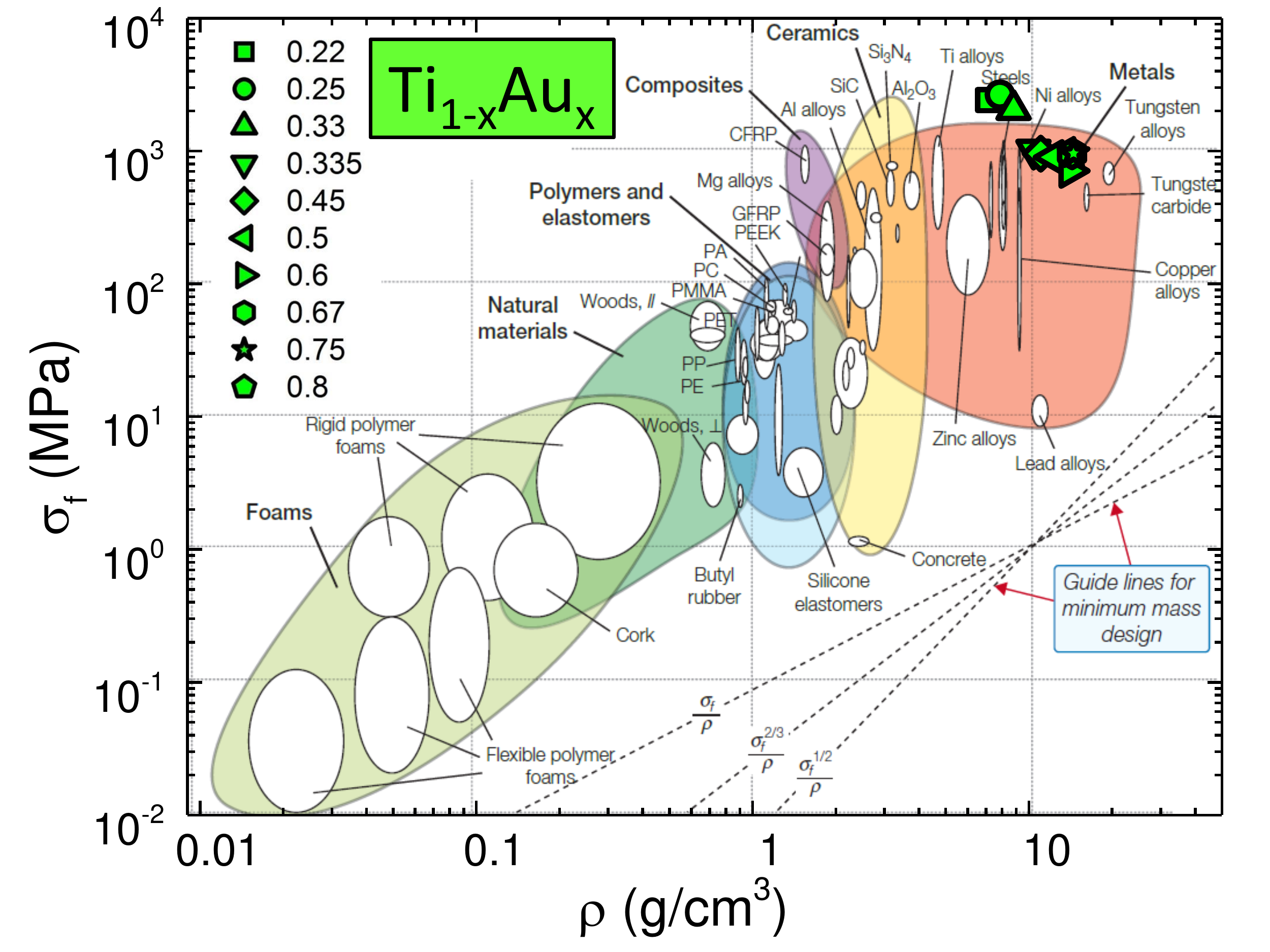}
\caption{Ashby diagram with strength $\sigma_f$ \textit{vs.} mass density $\rho$ for different families of materials \cite{Ashby_2011}, along with Ti$_{1-x}$Au$_x$ alloys (current study, green symbols).}
\label{Ashby}
\end{figure}

Since the mass density $\rho$ of Au is lowered by Ti dilution, as shown in Fig. \ref{Density}, Ti$_{1-x}$Au$_x$ alloys are even more desirable for biomedical applications which usually employ the denser Au, given the decreased density $\rho$ and increased hardness $H_V$, yielding stronger and lighter components with increasing $x$ in Ti$_{1-x}$Au$_x$. Melting the Ti$_{1-x}$Au$_x$ in Al$_2$O$_3$ crucibles resulted in the intermetalic alloy coating the walls of the ceramic container, making the mechanical removal of the coating impossible. This can reduce both the weight and cost of medical components, by using Ti$_{1-x}$Au$_x$ as a coating for a ceramic part. Moreover, the previously observed radio opacity of Ti$_{1-x}$Au$_x$ ($0.45 \leq x \leq 0.55$) alloys \cite{Boylan_2009, Boylan_2008} allows for radiographic distinction between implants and live tissue.

The melting temperatures $T_m$ of the Ti$_{1-x}$Au$_x$ alloys, extracted from the Ti-Au phase diagram \cite{ASM_2007}, are plotted in Fig. \ref{Density} and summarized in Table \ref{HardnessT}. The values are higher than those of both Ti and Au, which preliminarily classifies Ti$_{1-x}$Au$_x$ alloys as refractory metals \cite{ASTM_1986}. Detailed wear analysis is necessary to prove the refractory nature of the Ti$_{1-x}$Au$_x$ alloys. The potential applications include high-temperature engineering structures, heat sinks, electronics, aerospace, metal cutting and forming, as well as the tooling industry \cite{Smallwood_1983}. 

Preliminary strength estimates, determined from hardness values, place Ti$_{1-x}$Au$_x$ alloys (green symbols, Fig. \ref{Ashby}) just above the engineering metals region of the strength-mass density Ashby diagram. While other metallic alloys such as tungsten carbide, high-carbon steels, and nickel alloys show similar hardness values, they are often not desirable for medical applications due to high density and high toxicity.

\begin{table*}[hbtp]
\begin{tiny}
\renewcommand{\arraystretch}{1.25}
\renewcommand{\tabcolsep}{0.2cm}
\caption{\label{HardnessT} Summary of parameters for as-cast Ti$_{1-x}$Au$_x$ alloys.}
\begin{center}
\begin{tabular}{c|c|c|c|c|c|c}
	\hline \hline

\multirow{2}{*}{$x$}&Melting temperature $T_{m}$\cite{ASM_2007}&Mass density $\rho$&As-cast hardness $H_V$&As-cast strength $\sigma_f$&TEM                       &X-ray      \\
                    &($^\circ$C)                               &(g/cm$^3$)         &(GPa)                 &(GPa)                      &composition               &composition\\ \hline \hline
0                   &882                                       &4.45               &$2.62 \pm 0.22$       &$0.89 \pm 0.07$            &--                        &--         \\ \hline
0.22                &1367                                      &7.14               &$7.11 \pm 0.12$       &$2.42 \pm 0.04$            &Ti$_3$Au + $\alpha$Ti     &Ti$_3$Au   \\ \hline
0.25                &1395                                      &7.87               &$7.81 \pm 0.21$       &$2.65 \pm 0.07$            &                          &Ti$_3$Au   \\ \hline
0.33                &1310                                      &8.81               &$6.15 \pm 0.44$       &$2.05 \pm 0.1$             &TiAu + TiAu$_2$           &           \\ \hline
0.45                &1470                                      &10.39              &$2.99 \pm 0.40$       &$1.02 \pm 0.13$            &                          &           \\ \hline
0.50                &1495                                      &11.05              &$2.34 \pm 0.24$       &$0.80 \pm 0.08$            &$\beta$TiAu + Ti          &TiAu       \\ \hline
0.60                &1385                                      &12.25              &$2.64 \pm 0.24$       &$0.90 \pm 0.08$            &TiAu + TiAu$_2$ + TiAu$_4$&TiAu$_2$   \\ \hline
0.67                &1455                                      &14.09              &$2.09 \pm 0.05$       &$0.71 \pm 0.02$            &                          &TiAu$_2$   \\ \hline
0.75                &1400                                      &14.35              &$2.81 \pm 0.30$       &$0.95 \pm 0.10$            &                          &           \\ \hline
0.80                &1315                                      &14.52              &$2.83 \pm 0.08$       &$0.96 \pm 0.03$            &                          &TiAu$_4$   \\ \hline
1.00                &1064                                      &17.65              &$0.41 \pm 0.04$       &$0.14 \pm 0.01$            &--                        &--        \\ \hline \hline

\end{tabular}
\end{center}
\end{tiny}
\end{table*}

Given that Ti alloys are frequently heat-treated to improve both hardness and ductility, annealing studies were carried out for the Ti-Au system. However, annealing using two different recipes - for 7 days at 900$^\circ$C, or for several hours at 0.5$T_m$ and 0.3$T_m$ (similar to what has been done for other Ti-base alloys \cite{Rajan_1994}) resulted in minimal changes in the hardness compared to the as-cast samples, which might be caused by variation in microstructure homogeneity which can mask the true annealing effects.

\section{IV. Conclusions}

A series of Ti$_{1-x}$Au$_x$ alloys ($0.22 \leq x \leq 0.8$) has been investigated due to their extreme hardness values, elevated melting temperatures (compared to those of constituent elements), reduced density compared to pure Au, bulk metallicity, high biocompatibility and radio opacity. These properties make medical applications especially favorable with examples including replacement parts and components (both permanent and temporary), dental prosthetics and implants. The ability to adhere to ceramic components along with osseointegration are also beneficial, as they are able to reduce component weight and cost. Possible additional applications include circuit wires, hard coatings for tools and other medical equipment, drill head bits, as well as sporting goods. 

\section{Acknowledgements}

The work at Rice was supported by NSF DMR 0847681 (E.M. and E.S.).


\end{document}